\documentclass{jaa}
\usepackage{natbib}
\usepackage{graphicx}
\begin{document}\sloppy

\title{Swift J1753.5-0127:\\ Understanding the accretion geometry through frequency resolved spectroscopy}


\author{Blessy E. Baby\textsuperscript{1*} \and M. C. Ramadevi\textsuperscript{2}}
\affilOne{\textsuperscript{1}Department of Physics, University of Calicut, Malappuram 673635, Kerala, India.\\}
\affilTwo{\textsuperscript{2}Space Astronomy Group, ISITE campus, U. R. Rao Satellite Centre, Karthik Nagar, Bangalore 560037, Karnataka, India}


\twocolumn[{

\maketitle

\corres{blessy.elizabeth65@gmail.com}

\msinfo{1 January 2020}{1 January 2020}

\begin{abstract}
The Black Hole Binary source Swift J1753.5-0127 remained in outburst for $\sim$ 12 years from May 2005 to April 2017. For most part of the outburst, the source remained in the Low Hard State (LHS) displaying transitions to softer states only towards the end of the outburst for short periods of time. Quasi periodic Oscillations (QPOs) were observed in the Power Density Spectrum (PDS) only during the decay. A soft thermal component was required to model the spectrum in LHS, which does not conform to the generally accepted disc truncation theory. In this work, we attempt to obtain a clearer picture of the accretion disc geometry by studying the QPO variability using frequency resolved spectroscopy (FRS). We obtain the QPO rms spectrum of the source during the bright-hard state and model it with physical components. We find that the QPO rms spectrum can be described only by a Comptonization component with no contribution from the thermal disc. This indicates that the variability observed in the PDS originates in the Comptonization component and the evolution of the QPOs is likely to be a result of localization of the variabilities to different radii of the hot inner flow rather than disc truncation. The minimal variation in disc parameters also points to the existence of a stable disc throughout the outburst. 
\end{abstract}


}]


\doinum{12.3456/s78910-011-012-3}
\artcitid{\#\#\#\#}
\volnum{000}
\year{0000}
\pgrange{1--8}
\setcounter{page}{1}
\lp{8}

\section{Introduction}

Black Hole Binaries (BHBs) are a subset of low mass X-ray binaries (XRBs), which consist of a black hole accreting matter from a low mass star. Most of the BHBs exist as transients, where the source brightens significantly for a few weeks to months, or sometimes years, and then decays to quiescence. These outbursts are generally explained by the disc-instability model \citep{Lasota2001,Dubus2001} where accretion of matter onto the disc triggers the transformation of a cool, neutral disc to a hot, ionized one. This results in the viscous transfer of matter through the disc, which then falls onto the compact object \citep[see][for reviews]{Remillard2006,Done2007}. 

The energy spectra obtained from these sources is generally attributed to a multicolour disc \citep{Shakura1973} and the Comptonization of disc photons by hot, energetic plasma electrons \citep{Sunyaev1980,Narayan1996}, generally modelled as a powerlaw component. Varying contributions from these two components results in the classification of the outbursts into different states. In a canonical outburst, the source undergoes a transition from Low/Hard State (LHS) to High/Soft State (HSS), through Hard Intermediate (HIMS) and Soft Intermediate states (SIMS) before going back to the LHS at the end of the outburst \citep[][etc.]{Homan2001,Homan2005,Nandi2012,Baby2020}. The disc increases in prominence while the powerlaw component decreases, and sometimes disappears, as the source moves from LHS to HSS. The variabilities in the source can be studied by performing the Fourier transform of the lightcurve, which gives the PDS. The PDS typically shows weak red noise in the HSS, Band Limited Noise (BLN) or peaked Lorentzian features called Quasi-Periodic Oscillations in the intermediate states and LHS \citep[][etc.]{Casella2004, Remillard2006}. The rms variability of BHBs varies from $>$ 20\% in the intermediate and hard state to $<$ 5\% in the disc dominant states \citep{Remillard2006,Baby2020}. Type C QPOs are dominant in the PDS during LHS and HIMS, whereas Type A/B type QPOs are sometimes seen in SIMS. Soft states are marked by an absence of QPOs. \citep{Belloni2002,Casella2004,Remillard2006,Belloni2014}. The Hardness-Intensity Diagram (HID) can also serve as a marker to identify state transitions. It is defined as the plot between ratio of fluxes in hard to soft band and the total flux from the source. The HID starts from LHS at the lower right and passes through HIMS, SIMS, HSS and then returns to LHS, literally tracing out a `q' in the plot \citep[][and references therein]{Homan2001,Homan2005,Nandi2012,Sreehari2019}. Deviation from this standard picture is seen for a few sources where the source remains in either the hard state or the soft state for most of the duration of the outburst. Such outbursts are termed as `failed' outbursts. The correlated spectral and timing studies provide us an opportunity to study the underlying physics behind such outbursts and the apparent similarities/dissimilarities among them. 

Swift J1753.5-0127 (hereafter J1753) is a BHB discovered by \textit{Burst Alert Telescope (BAT)} \citep{Barthelmy2005} onboard \textit{Neil Gehrels Swift} in May 2005 \citep{Palmer2005}, which remained in outburst for a period of $\sim$ 12 years from May 2005 to April 2017 \citep{Shaw2016c}. Re-brightening or mini-outbursts were also reported from January to April 2017 \citep{Bright2017,Tomsick2017,Bernardini2017}. A lower limit of mass of the compact object is placed at 7.4  $M_{\odot}$ using the double-peaked hydrogen emission lines in the optical spectrum \citep{Shaw2016b}. The source is one of the few systems with an orbital period $<$ 5 hr ($P_{orb}$ = 3.2 h) \citep{Zurita2008}. Although the source is considered to be at an inclination $\leq$ 55$^{\circ}$ \citep{Froning2014,Shaw2016a}, a precise measurement of inclination and distance is not available. J1753 reached a peak flux of 200 mCrab within a week of its detection, as observed by \textit{Rossi X-ray Timing Explorer (RXTE)} and then decreased within a few months to 20 mCrab. It remained constant at this flux for most of the duration of the outburst with a few intermittent bursts where transition to HIMS or soft states was seen and reported using quasi-simultaneous data from \textit{XMM-Newton} and \textit{NuSTAR} \citep{Shaw2016a}. However, during the Fast Rise and Exponential Decay (FRED) profile observed at the beginning of the outburst, the source was found to remain in the LHS, thus being classified as a Low Hard X-ray transient (LHXT) initially \citep{Ramadevi2007, Zhang2007}. The soft excess in the spectra was attributed to the presence of disc close to the compact object even in the LHS in a few studies \citep[for e.g.,][]{Ramadevi2007}. Alternatively, \cite{Shaw2019} find the physical properties of the source to be compatible with the presence of a fully irradiated disc truncated at large radii by investigating the emission mechanisms during mini-outbursts.

 Frequency resolved spectroscopy (FRS) \citep{Revnivtsev1999, Gilfanov2000,Sobolewska2006} can break this apparent degeneracy of spectral modelling. In this work, we obtain the QPO spectra using FRS and fit it with physical models. This can help us understand the physical origin of the QPO and also provide some insight as to the mechanism behind its evolution. QPO origin theories can be divided into two broad categories. The first is due to instabilities in the transition layer where Keplerian flow from accretion disc meets the sub-Keplerian flow close to compact object \citep{Titarchuk2004}. The second one is due to geometrical effects i.e., relativistic precession of the radially extended hot inner flow \citep{Ingram2009}. In either case, studying the evolution of QPO spectrum with respect to the time-averaged spectrum, can provide a better picture of the accretion geometry of the source. QPOs are observed throughout the decay phase of the outburst of the source J1753. Here, we model the QPO spectrum in an attempt to understand its physical origin and thereby comment on the geometry of accretion. 

In Section 2, we detail the observations and the data reduction methods used to obtain time averaged spectra and FRS. We compare the evolution of the QPO spectra with the time-averaged spectra in Section 3. Finally, in Section 4, we discuss the origin of the soft excess in the time-averaged spectra in the context of our findings  and present the conclusions.
\begin{figure*}  
\centering
\includegraphics[scale=0.5]{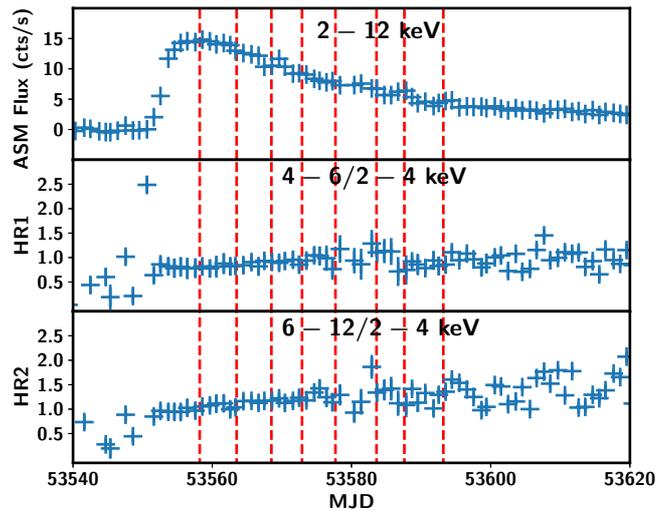}

\caption{\textit{ASM} lightcurve of the outburst from 19 June 2005 to 09 September 2005 in the energy band $2-12$ keV is plotted in the top Panel. Hardness ratios HR1 and HR2 are defined as ratio of counts in B band to A band and C band to A band respectively and their variation with time is shown in Panels 2 and 3. The vertical red dashed lines represent the observations chosen for analysis.}
\label{fig:lc}
\end{figure*}
\section{Observations and Data Reduction}

\textit{RXTE} observed the source for a total of 344 times from the beginning of its outburst on 2 July 2005 to 29 November 2011. Of these 344 observations, type C QPOs were observed in 46 observations, most of them at the beginning of the outburst. The source transitioned to HIMS in 2016. However, QPOs were not observed ,and therefore, QPO spectra can only be studied using observations in \textit{RXTE} era. To perform frequency resolved spectroscopy, we require data with simultaneously good spectral and temporal resolution. The top panel of Figure \ref{fig:lc} shows the \textit{ASM} lightcurve of the source from 19 June to 09 September 2005. Lightcurves are also obtained in A : $2-4$, B : $4-6$ and C : $6-12$ keV energy bands. The hardness ratios, HR1 and HR2, are obtained as ratio of counts in B to A band and C to A band respectively. These are plotted in the middle and bottom panels of Figure \ref{fig:lc}.  Most of the observations performed with the proposal ID P91423 could be used to perform FRS using both event and binned data. We therefore chose eight observations (out of 30 observations where QPOs are reported in P91423), to generate QPO spectrum. The observations are chosen such that they are separated by a time interval of $\sim$ 5 days.  These are marked as dashed vertical lines in Figure \ref{fig:lc}. We use Standard-2 files to generate the time averaged spectrum from \textit{PCA} data in the $3-25$ keV range. Event mode data files are used to generate the dead-time corrected lightcurves with a bin time of 8 ms. We use only PCU2 data for both the spectral and light curve extraction. Standard procedures are used to extract \textit{HEXTE} data to obtain the energy spectrum from $25-150$ keV. Only Cluster B is used as Cluster A had stopped rocking at that point.

PDS were generated for all the eight observations using a bin time of 8 ms and 8192 bins per intervals. The PDS for the first and last of the observations considered are plotted in Figure \ref{fig:pds}. 

To generate frequency resolved spectra of the QPOs, we chose the event and binned mode data files with a timing resolution greater than 5 ms. We check for the data grouping using TEVTB2 keyword in the header file to ensure that channels corresponding to $8-13$ keV are not binned together. We use \textit{stingray} module in \texttt{python} to generate the PDS in each available energy channel \citep{Huppen2019} with rms-squared normalization \citep{Belloni1990,Miyamoto1992}. The rms values were then used to obtain counts using the relation 
\begin{equation}
S=R \times \sqrt{P \delta f}
\end{equation}
where $R$ is the count rate, $P$ is the summed power over the QPO frequency range and $\delta f = 1/T$ where $T$ is the length of each segment. $P$ is obtained by modelling the QPO in each channel and integrating the counts rather than obtaining the sum of powers, which provides a better estimate. The PDS data with information on the channel, the count rate obtained using the above method (S) and the count rate error are saved into a text file. This text file is then converted to \texttt{XSPEC} readable file using \texttt{ascii2pha} command. Dead time correction and background subtraction is also performed. This file can then be loaded into \texttt{XSPEC} with the response files generated for Std2f files and fit with available models. 

\section{Spectral and Temporal Analysis}
\subsection{Temporal Analysis}
\begin{figure}[h!]
\includegraphics[width=\columnwidth]{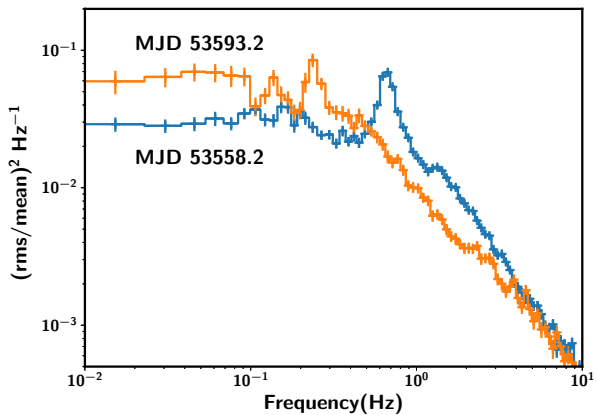}
\caption{PDS of first (MJD 53558.2) and eighth observations (MJD 53593.2) are plotted in blue and orange markers respectively. The centroid frequency of the QPO decreases from first to last observation.}
 \label{fig:pds}
\end{figure}

PDS were generated for all the eight observations using a bin time of 8 ms and 8192 bins per interval. A strong QPO is observed in all the cases varying from $0.2-0.7$ Hz. The QPO frequency decreases with time as seen in Fig. \ref{fig:pds}. A harmonic is also seen from $1.2-1.4$ Hz at the beginning which later disappears. A sub-harmonic is seen in the last observation on MJD 53593.2 at $\sim$ 0.13 Hz. 
\subsection{Time averaged spectra}
\begin{table*}
\begin{center}
\tabularfont
\caption{Fit parameters using the model \texttt{TBabs(diskbb+gaussian+nthcomp)}}\label{tab:specpar} 
\begin{tabular}{c|c|c|c|c@{\hspace{2pt}}|c@{\hspace{2pt}}|c@{\hspace{2pt}}|c@{\hspace{2pt}}|c@{\hspace{2pt}}|c|c}
\topline
ObsId & MJD &$T_{in}$ &$\Gamma$&$kT_{e}$ &$E_{line}$ &Eq. width & $\sigma$ & $N_{gauss}^{a}$ & $F_{3-25}^{b}$ & $\chi^{2}$/dof\\
& &(keV)& & (keV)& (keV)& (eV)& & & erg cm$^{-2}$ s$^{-1}$ & 
\\\midline
91423-01-01-00& 53558.2 & 0.84$^{+0.03}_{-0.02}$ & 1.69$^{+0.01}_{-0.01}$& 54$^{+5}_{-5}$ & 6.2$^{l}$ & 76$\pm$3 & $0.9\pm0.3$ & $4.3\pm0.2$ & $8.39\pm0.08$ & 72/80\\
QPO& & & 1.64$^{+0.04}_{-0.04}$& & & & & & & \\
91423-01-02-05& 53563.5 & 0.88$^{+0.04}_{-0.03}$ & 1.68$^{+0.01}_{-0.01}$& 54$^{+7}_{-10}$ & 6.9$^{l}$ & 74$\pm$4 & $1.0\pm0.5$ & $3.8\pm0.9$ & $7.43\pm0.02$ & 66/80\\
QPO& & & 1.1$^{+0.5}_{-0.3}$& & & & & & & \\
91423-01-03-02& 53568.5 & 1.01$^{+0.03}_{-0.03}$ & 1.65$^{+0.01}_{-0.01}$& 55$^{+7}_{-10}$ & 6.9$^{l}$ & 47$\pm$2 & $0.6\pm0.4$ & $1.6\pm0.4$ & $6.36\pm0.02$ & 51/80\\
QPO& & & 1.0$^{+0.4}_{-0.2}$& & & & & & & \\
91423-01-03-06& 53572.9 & 0.96$^{+0.04}_{-0.03}$ & 1.64$^{+0.01}_{-0.01}$& $63^{+13}_{-24}$ & 6.9$^{l}$ & $40\pm6$ & $0.6\pm0.4$ & $1.3\pm0.4$ & $5.74\pm0.01$ & 66/80\\
QPO& & & 1.1$^{+0.2}_{-0.7}$& & & & & & & \\
91423-01-04-04& 53577.7 & 0.99$^{+0.03}_{-0.03}$ & 1.62$^{+0.01}_{-0.01}$& 49$^{+6}_{-8}$ & 6.9$^{l}$ & 67$\pm$8 & $1.1\pm0.5$ & $2.1\pm0.4$ & $4.96\pm0.01$ & 79/80\\
QPO& & & 1.1$^{+0.2}_{-0.3}$& & & & & & & \\
91423-01-05-01& 53583.6 & 0.98$^{+0.02}_{-0.04}$ & 1.61$^{+0.01}_{-0.01}$& 59$^{+8}_{-14}$ & 6.2$^{l}$ & 69$\pm$2 & $0.9\pm0.5$ & $1.7\pm0.1$  & $4.14\pm0.08$ & 57/80\\
QPO& & & 1.1$^{+0.1}_{-0.1}$& & & & & & & \\
91423-01-06-00& 53587.6 & 0.92$^{+0.05}_{-0.04}$ & 1.59$^{+0.01}_{-0.01}$& 54$^{+8}_{-12}$ & 6.2$^{l}$ & 100$\pm3$ & $1.0\pm0.2$ & $2.2\pm0.3$ & $3.58\pm0.01$ & 61/80\\
QPO& & & 1.0$^{+0.2}_{-0.5}$& & & & & & & \\
91423-01-06-03 & 53593.2 & 0.84$^{+0.06}_{-0.04}$ & 1.59$^{+0.01}_{-0.01}$& 82$^{+32}_{-22}$ & 6.2$^{l}$ & 58$\pm$3 & $0.6\pm0.4$ & $1.6\pm0.5$ & $2.95\pm0.02$ & 91/80\\
QPO& & & 1.26$^{+0.07}_{-0.06}$& & & & & & &  \\
\hline
\end{tabular}
\end{center}
\tablenotes{$^{l}-$upper limit}
\tablenotes{$^{a}-$in units of $10^{-3}$}
\tablenotes{$^{b}-$Flux is in units of $10^{-9}$ erg cm$^{-2}$ s$^{-1}$}

\end{table*}

\begin{figure}[h!]
\includegraphics[scale=0.55]{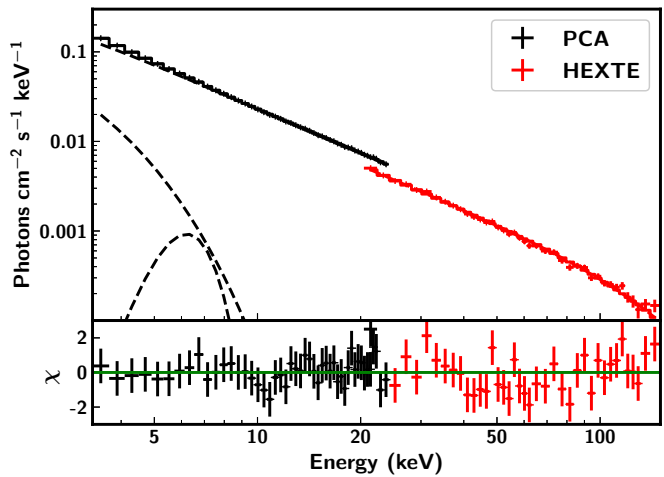}
\caption{The above figure shows the broadband spectrum of Obsid 91423-01-02-05 fit with Model \textit{TBabs(diskbb+gaussian+nthcomp).}}
 \label{fig:spec}
\end{figure}
We obtained the time averaged spectra for the eight observations in consideration with \textit{PCA} and \textit{HEXTE} data from 3-25 keV and 20-150 keV respectively. We first fit the broadband spectrum with phenomenological models. Considering that the source is in the LHS, we fit the spectrum using a power law model along with \textit{TBabs} \citep{Wilms2000} to account for the interstellar absorption. The abundance is set at \texttt{wilm} with \texttt{vern} cross-section. The $N_{H}$ value could not be constrained and was frozen at $2 \times 10^{21}$ cm$^{-2}$ \citep{Cadolle2007}. A simple power law could not fit the spectrum and hence was replaced by a cut-off powerlaw. However, significant residuals were seen at lower energy, which required the addition of \textit{diskbb} \citep{Mitsuda1984,Makishima1986} model. A \textit{gaussian} was included to account for residuals seen around 6 keV. This resulted in good fits with $\chi^{2}_{red}$ $=$ 81/80. A normalization constant was also included to account for calibration difference in both the instruments. To obtain a better understanding of the geometry of the system, we replace the \textit{powerlaw} model with \textit{nthcomp} \citep{Zdziarski1996,Zycki1999}, which considers the Comptonization of the disc photons by hotter electrons. The final model used was \textit{TBabs (diskbb+gaussian+nthcomp)}. The unfolded spectrum of ObsId 91423-01-02-05 is shown in Figure \ref{fig:spec} as an example. \textit{PCA} and \textit{HEXTE} data are represented by black and red markers respectively. The fit parameters are given in Table \ref{tab:specpar}. The temperature of the seed photons ($kT_{bb}$) was tied to the disc temperature where we consider the seed photons to originate from the disc. As the source was found to be in a hard state, we also attempt to replace \textit{diskbb} with a softer Comptonization component like the \textit{compTT} model. Although the fits are comparable with those obtained with the final model ($\chi_{red}^2$=73/77), the parameters of the \textit{compTT} model could not be constrained. As no new information is obtained by the addition of this model, we continue to use the previous one and fit all the available observations with \textit{TBabs (diskbb+gaussian+nthcomp)} model as described above. Here, we would like to emphasize that the normalization of the disc component could not be constrained in a few cases. This is probably due to the lack of low energy spectra ($0.1-3$ keV), where the thermal disc is likely to be prominent.


\subsection{Energy spectra of QPO}

\begin{figure}[h!]
\includegraphics[scale=0.55]{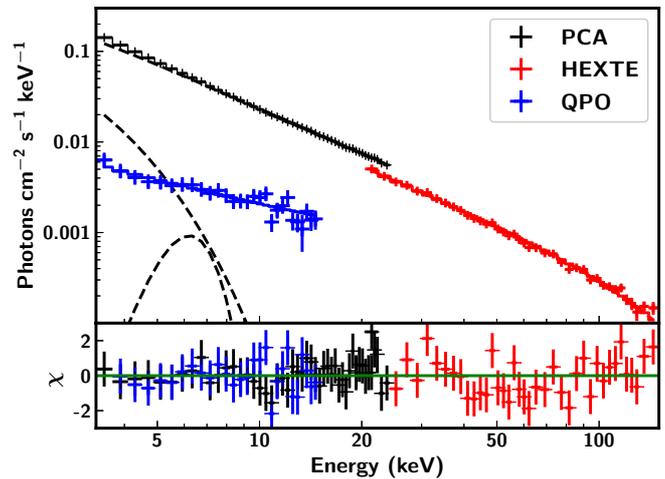}
\caption{The time averaged spectrum is plotted with QPO rms spectrum for the ObsID: 91423-01-02-05. The QPO rms spectrum is fit with an absorbed Comptonization model and is scaled up for clarity.}
 \label{fig:qpo_with_tot}
\end{figure}

Low frequency QPOs were observed in the PDS throughout the exponential decay phase of the outburst. Here we obtain the energy spectrum corresponding to the variability in the frequency range where the QPO is observed. We follow \cite{Sobolewska2006,Axelsson2014,Axelsson2016} to obtain the QPO rms spectra for all the eight observations considered. We could fit the spectrum with an absorbed Comptonization component in all the cases. The final model used was \textit{TBabs*nthComp}. There was no signature of a soft disc or a reflection component in the QPO rms spectrum. The temperature of the input seed photons could not be constrained due to statistical limitations. Therefore, we freeze the temperature to the one obtained from fitting the corresponding time-averaged spectrum. The origin of the Comptonized photons is assumed to be from the disc. The electron plasma temperature also could not be constrained for a few observations as shown in Table \ref{tab:specpar}. Therefore, it was also frozen to the values obtained using fits from time-averaged spectra. Figure \ref{fig:qpo_with_tot} shows the fitted time-averaged spectrum along with the QPO rms spectrum for ObsId 91423-01-02-05. \textit{PCA} and \textit{HEXTE} data are marked as black and red crosses respectively whereas QPO rms spectrum is shown in blue. The corresponding residuals are shown in the bottom panel.

\section{Results}
\subsection{Outburst profile and HID}

J1753 was in outburst from May 2005 to April 2017. The fast rise and exponential decay (FRED) profile was observed at the beginning of the outburst which lasted for a few months. The top panel of Figure \ref{fig:lc} shows the \textit{ASM} lightcurve of the source in $2-12$ keV range. The source was found to be in the LHS for the entire duration of the FRED profile. The eight observations chosen for study of the source with FRS are represented by red dashed lines. Hardness ratios HR1 and HR2, as defined earlier, are plotted in the middle and bottom panels. A sudden increase in HR1 is seen before the FRED profile. However, pointed observations for the date do not exist. Other than this isolated spike in HR1 and a slight dip in HR2 at the beginning of the outburst, there is no clear evidence pointing to a probable state transition. HR2 seems to increase slightly beyond MJD 53600 as seen from panel 3. Here again, the variation is too low and does not support a state transition.

\begin{figure}[h!]
	\includegraphics[width=\columnwidth]{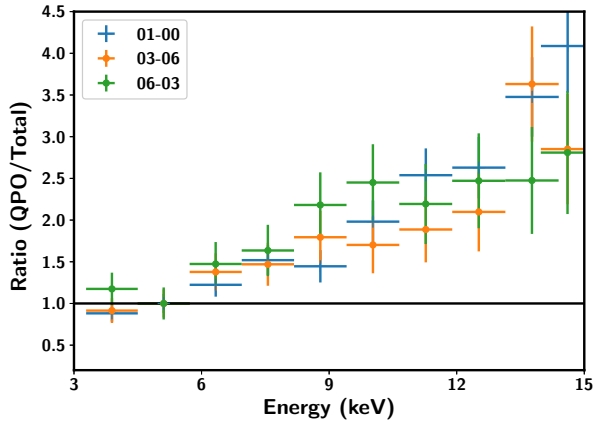}
	\caption{The ratio of the QPO rms spectrum to the total spectrum for three observations is shown in the figure. QPO spectra are always harder than the time averaged spectra. The data is rebinned and the counts are normalized to 1 at 5 keV for a clearer representation (see text for details).}
	\label{fig:qpo_ratio}
\end{figure}
\begin{figure}
\includegraphics[width=\columnwidth]{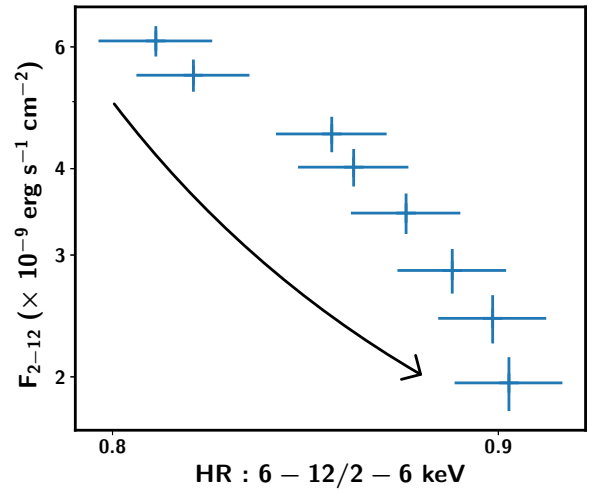}
\caption{HID obtained from \textit{RXTE/PCA} is shown in the figure. The hardness ratio decreases along with flux with time as indicated by the arrow. HR increases from 0.8 to 0.9 for the eight observations considered.}
\label{fig:hid}
\end{figure}

\subsection{Time-averaged spectra vs QPO spectra}

Results of fits with the model \textit{TBabs (diskbb+gaussian+nthcomp)} along with fits to QPO spectra are presented in Table \ref{tab:specpar}. The time-averaged spectrum and QPO spectrum obtained for ObsId 91423-01-02-05 are shown in Figures \ref{fig:spec} and \ref{fig:qpo_with_tot} as an example. As the source moves to the faint-hard state, the time-averaged spectra become harder while the disc temperature shows a slight increase from 0.84 to 1.01 keV within 10 days and again reduces to 0.84 keV. No discernible pattern is seen in the equivalent width of the \textit{Gaussian}. The fit results for the QPO spectra are also shown in Table \ref{tab:specpar} for comparison. The photon indices obtained for the QPO rms spectrum are harder than those for the time-averaged spectra. However, the errors on the values are statistically limited. To further probe the evolution of QPO spectra with respect to the time-averaged spectra, we plot the ratio of QPO rms spectrum to the total spectrum in Figure \ref{fig:qpo_ratio}. The data is grouped to contain a minimum of 5 counts per bin. The counts are  normalized to 1 at 5 keV and for clarity \citep{Axelsson2016}. Although Figure \ref{fig:qpo_ratio} suggests that the QPO rms spectrum becomes harder as the source moves toward the LHS, not much evolution is seen in the QPO spectra between observations taken 40 days apart. It is also observed that the QPO spectra are harder than the time-averaged spectra in all cases.

\section{Discussion and Conclusion}

J1753 has exhibited peculiar behaviour from the time of its detection, including an unusually long-period outburst for a source with small orbital period ($P_{orb}=3.4$ hrs). Observation of mini-outbursts is also reported towards the end of the main outburst \citep{Shaw2016b}. Although classified to be in the hard state, a soft excess at the lower energy range was observed in most of the cases supporting the existence of disc even in the LHS. We therefore perform variability studies of the source to comment on its dynamic nature. FRS technique allows us to isolate and localize regions of fast variability and study the spectrum in detail. The strong Type-C QPOs observed in the PDS of the source serve as an excellent candidate to perform FRS and localize the source of variability. 

A canonical outburst typically follows LHS$\rightarrow$ HIMS $\rightarrow$ SIMS$\rightarrow$ HSS$\rightarrow$ SIMS $\rightarrow$HIMS $\rightarrow$ LHS path tracing out a `q'- shape in the HID. However, state transitions were not reported for the source during its main outburst as it remained only in the LHS. Hence, this outburst is termed as a `failed' outburst. To trace out the precise path of the source on the HID, we obtain the model fitted flux using \textit{RXTE/PCA} data in $2-6$ keV and $6-12$ keV energy bands. This is shown in Figure \ref{fig:hid}. HR decreases systematically along with flux as the source moves to harder state.  As seen in Figure \ref{fig:hid}, the source moves from the top left to bottom right during the decay in the main outburst with very less variation in the hardness ratio. It is clear that the source does not follow the `q'-shaped profile. A few soft X-ray transients seem to follow a `c' shaped profile in the HID like 4U 1630-472 \citep{Capitanio2015,Baby2020} and MAXI J0637-430 \citep{Baby2021}. Although, data presented here is insufficient to draw such a conclusion, it is possible that the source could evolve further to follow a similar profile here. It is also likely that the source could have transitioned to a soft state briefly (lasting for a few hours), which was missed by the \textit{RXTE/PCA} instrument. Fast transitions have also been reported at the beginning of the outburst of 4U 1630-472 from the hard to soft state using \textit{SXT} onboard \textit{AstroSat} \citep{Baby2020}, after which it follows a similar profile as in Figure \ref{fig:hid}. Lack of observations at the beginning of the outburst make it difficult to ascertain if such a transition occurred in this source. However, the similarities in the evolutionary track of the HID followed by soft and hard X-ray transients are suggestive of a common mechanism at play. Detailed comments on this aspect require a grouping studies, considering all such sources, which is beyond the scope of this work.

Similar to earlier studies \citep[][etc.]{Ramadevi2007,Zhang2007,Shaw2016b,Kajava2016,Shaw2019}, we also find that a disc component is required to model the spectra in the hard state. This was explained either by a disc extending close to the innermost stable circular orbit (ISCO) \citep{Ramadevi2007} or an irradiated disc truncated at large radii where a large fraction of X-rays is reprocessed in the outer disc and the source of irradiation is produced in the corona within the truncation radius \citep{Shaw2019}. \cite{Kajava2016} suggest the source of seed photons to be from synchrotron self Comptonization below a critical flux limit which conforms to the truncated disc scenario but fails to explain the constant soft excess seen in the spectra. We observe that the inclusion of a soft Comptonization component instead of a thermal disc can also fit the spectra satisfactorily. However, the parameters corresponding to this component could not be constrained. This makes it difficult to comment on the origin of the soft Comptonization component. Nevertheless, existence of an inhomogeneous corona cannot be ruled out. Changes in the disc temperature and equivalent width of reflection line are too small to suggest a significant change in the accretion disc geometry (see Table \ref{tab:specpar}). This is consistent with a physical scenario where the accretion disc remains stable throughout the outburst with spectral and temporal variations driven by changes only in the hot inner flow. Also, as the disc parameters could not be constrained in some cases, we would like to reiterate that although the presence of disc is indisputable, the parameters obtained from the thermal disc component could be unreliable and further inferences based on these values is unwarranted.   
 
FRS studies of the source can help in determining the cause for variability observed in the PDS. In localizing the origin of the variability, we can gain further information on the accretion geometry of the source. The PDS is characterized by a band limited noise (BLN) and a QPO (see Figure \ref{fig:pds}) in all cases. The QPO rms spectrum is found to be slightly harder than the time averaged spectra and does not require the addition of a thermal disc component (see Table \ref{tab:specpar} and Figure \ref{fig:qpo_with_tot}). The Comptonization component used to model the QPO rms spectrum seems to follow the main Comptonization. Thus it follows that the origin of the variability that produces the QPO is seeped in the origin of the Comptonization component although at smaller radii \citep{Axelsson2018}. This suggests that the variabilities are localized to different radii in the inner flow but are not likely to be associated with the disc. 

The above results suggest that the disc is present and stable throughout the outburst with the temporal variations likely to originate in the hot inner flow. The change in QPO frequency has long been associated with a receding disc in the low hard state. Here we find that the evolution in the QPO is independent of the disc and is more likely to be a manifestation of the variations in the Comptonization component or localization of the variability to different radii in the hot inner flow. The mini-outburst reported towards the end of the main outburst also complicates the accretion picture as it requires the existence of a hot disc towards the end of the main outburst \citep{Zhang2019}. Nevertheless, the lack of variability originating from the disc in the PDS favours a stable disc rather than a dynamic one.

In this work, we present the spectral modelling of the time-averaged spectra and FRS of the BHB Swift J1753.5-0127 using \textit{RXTE} data. Using FRS studies, we localize the origin of the QPO to the hot inner flow rather than the disc. The disc does not seem to contribute to the variability seen in the source, suggesting that it is either absent or stable. The soft excess seen in the time-averaged spectra and the detection of a mini-outburst suggests the existence of a stable disc at the end of the main outburst, which agrees with the above analysis. Application of FRS technique to similar sources would help to provide a better picture of the accretion disc geometry of BHBs.








\section*{Acknowledgements}
The authors thank the reviewer for his/her valuable comments which greatly improved the quality of this paper.
MCR thanks GH, SAG, DD, PDMSA and Director, URSC for encouragement and continuous support to carry out this research. This work has made use of data from HEASARC.
\vspace{-1em}


\bibliography{ref_swift}




\end{document}